\documentclass[aps,prl,twocolumn,showpacs,superscriptaddress]{revtex4}
\usepackage{graphicx}
\usepackage{amsmath}
\usepackage{amssymb}

\begin{document}
\title{Proposed Bell Experiment with Genuine Energy-Time Entanglement}
\author{Ad\'{a}n Cabello}
\email{adan@us.es} \affiliation{Departamento de F\'{\i}sica Aplicada
II, Universidad de Sevilla, E-41012 Sevilla, Spain}
\author{Alessandro Rossi}
\affiliation{Dipartimento di Fisica della ``Sapienza''
Universit\`{a} di Roma, I-00185 Roma, Italy, and\\
Consorzio Nazionale Interuniversitario per le Scienze Fisiche della
Materia, I-00185 Roma, Italy}
\author{Giuseppe Vallone}
\affiliation{Centro Studi e Ricerche ``Enrico Fermi'', Via
Panisperna 89/A, Compendio del Viminale, Roma I-00184, Italy}
\affiliation{Dipartimento di Fisica della ``Sapienza''
Universit\`{a} di Roma, I-00185 Roma, Italy, and\\
Consorzio Nazionale Interuniversitario per le Scienze Fisiche della
Materia, I-00185 Roma, Italy}
\author{Francesco De Martini}
\affiliation{Dipartimento di Fisica della ``Sapienza''
Universit\`{a} di Roma, I-00185 Roma, Italy, and\\
Consorzio Nazionale Interuniversitario per le Scienze Fisiche della
Materia, I-00185 Roma, Italy}
\author{Paolo Mataloni}
\affiliation{Dipartimento di Fisica della ``Sapienza''
Universit\`{a} di Roma, I-00185 Roma, Italy, and\\
Consorzio Nazionale Interuniversitario per le Scienze Fisiche della
Materia, I-00185 Roma, Italy}
\date{\today}




\begin{abstract}
Franson's Bell experiment with energy-time entanglement [Phys. Rev.
Lett. {\bf 62}, 2205 (1989)] does not rule out all local hidden
variable models. This defect can be exploited to compromise the
security of Bell inequality-based quantum cryptography. We introduce
a novel Bell experiment using genuine energy-time entanglement,
based on a novel interferometer, which rules out all local hidden
variable models. The scheme is feasible with actual technology.
\end{abstract}


\maketitle


Two particles exhibit ``energy-time entanglement'' when they are
emitted at the same time in an energy-conserving process and the
essential uncertainty in the time of emission makes
undistinguishable two alternative paths that the particles can take.
Franson \cite{Franson89} proposed an experiment to demonstrate the
violation of local realism \cite{Bell64} using energy-time
entanglement, based on a formal violation of the Bell
Clauser-Horne-Shimony-Holt (CHSH) inequality \cite{CHSH69}. However,
Aerts {\em et al.} \cite{AKLZ99} showed that, even in the ideal case
of perfect preparation and perfect detection efficiency, there is a
local hidden variable (LHV) model that simulates the results
predicted by quantum mechanics for the experiment proposed by
Franson \cite{Franson89}. This model proves that ``the Franson
experiment does not and cannot violate local realism'' and that
``[t]he reported violations of local realism from Franson
experiments \cite{KVHNC90} have to be reexamined'' \cite{AKLZ99}.

Despite this fundamental deficiency, and despite that this defect
can be exploited to create a Trojan horse attack in Bell
inequality-based quantum cryptography \cite{Larsson02}, Franson-type
experiments have been extensively used for Bell tests and Bell
inequality-based quantum cryptography \cite{TBZG00}, have become
standard in quantum optics \cite{Paul04, GC08}, and an extended
belief is that ``the results of experiments with the Franson
experiment violate Bell's inequalities'' \cite{GC08}. This is
particularly surprising, given that recent research has emphasized
the fundamental role of a (loophole-free) violation of the Bell
inequalities in proving the device-independent security of key
distribution protocols \cite{Ekert91}, and in detecting entanglement
\cite{HGBL05}.

Polarization entanglement can be transformed into energy-time
entanglement \cite{Kwiat95}. However, to our knowledge, there is no
single experiment showing a violation of the Bell-CHSH inequality
using genuine energy-time entanglement (or ``time-bin entanglement''
\cite{BGTZ99}) that cannot be simulated by a LHV model. By
``genuine'' we mean not obtained by transforming a previous form of
entanglement, but created because the essential uncertainty in the
time of emission makes two alternative paths undistinguishable.

Because of the above reasons, a single experiment using energy-time
entanglement able to rule out all possible LHV models is of
particular interest. The aim of this Letter is to describe such an
experiment by means of a novel interferometric scheme. The main
purpose of the new scheme is not to compete with existing
interferometers used for quantum communication in terms of practical
usability, but to fix a fundamental defect common to all of them.

We will first describe the Franson Bell-CHSH experiment. Then, we
will introduce a LHV model reproducing any conceivable violation of
the Bell-CHSH inequality. The model underlines why a Franson-type
experiment does not and cannot be used to violate local realism.
Then, we will introduce a new two-photon energy-time Bell-CHSH
experiment that avoids these problems and can be used for a
conclusive Bell test.


\begin{figure}[b]
\centerline{\includegraphics[width=1.04 \columnwidth]{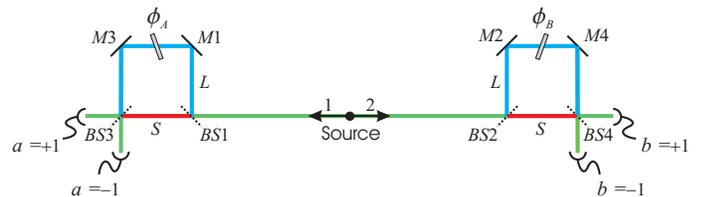}}
\caption{\label{Fig1} Generic setup of the Franson Bell experiment.}
\end{figure}


{\em The Franson Bell-CHSH experiment.---}The setup of a Franson
Bell-CHSH experiment is in Fig.~\ref{Fig1}. The source emits two
photons, photon $1$ to the left and photon $2$ to the right. Each of
them is fed into an unbalanced interferometer. $BS_i$ are beam
splitters and $M_i$ are perfect mirrors. There are two distant
observers, Alice on the left and Bob on the right. Alice randomly
chooses the phase of the phase shifter $\phi_A$ between $A_0$ and
$A_1$, and records the counts in each of her detectors (labeled
$a=+1$ and $a=-1$), the detection times, and the phase settings at
$t_D-t_I$, where $t_D$ is the detection time and $t_I$ is the time
the photon takes to reach the detector from the location of the
phase shifter $\phi_A$. Similarly, Bob chooses $\phi_B$ between
$B_0$ and $B_1$, and records the counts in each of his detectors
(labeled $b=+1$ and $b=-1$), the detection times, and the phase
settings. The setup must satisfy four requirements: (I) To have
two-photon interference, the emission of the two photons must be
simultaneous, the moment of emission unpredictable, and both
interferometers identical. If the detections of the two photons are
coincident, there is no information about whether both photons took
the short paths $S$ or both took the long paths $L$. A simultaneous
random emission is achieved in actual experiments by two methods,
both based on spontaneous parametric down conversion. In energy-time
experiments, a non-linear crystal is pumped continuously by a
monochromatic laser so the moment of emission is unpredictable in a
temporal window equal to the coherence time of the pump laser. In
time-bin experiments, a non-linear crystal is pumped by pulses
previously passing through an unbalanced interferometer, so it is
the uncertainty of which pulse, the earlier or the later, has caused
the emission what provokes the uncertainty in the emission time. In
both cases, the simultaneity of the emission is guaranteed by the
conservation of energy. (II) To prevent single-photon interference,
the difference between paths $L$ and $S$, i.e., twice the distance
between $BS1$ and $M1$, $\Delta {\cal L}=2 d(BS1,M1)$ (See
Fig.~\ref{Fig1}), must satisfy $\Delta {\cal L} > c t_{\rm coh}$,
where $c$ is the speed of light and $t_{\rm coh}$ is the coherence
time of the photons. (III) To make distinguishable those events
where one photon takes $S$ and the other takes $L$, $\Delta {\cal
L}$ must satisfy $\Delta {\cal L} > c \Delta t_{\rm coinc}$, where
$\Delta t_{\rm coinc}$ is the duration of the coincidence window.
(IV) To prevent that the local phase setting at one side can affect
the outcome at the other side, the local phase settings must
randomly switch ($\phi_A$ between $A_0$ and $A_1$, and $\phi_B$
between $B_0$ and $B_1$) with a frequency of the order $c/D$, where
$D=d({\rm Source},BS1)$.

The observers record all their data locally and then compare them.
If the detectors are perfect they find that
\begin{subequations}
\begin{align}
P(A_i=+1)=P(A_i=-1)=\frac{1}{2}, \label{Amarginal} \\
P(B_j=+1)=P(B_j=-1)=\frac{1}{2}, \label{Bmarginal}
\end{align}
\end{subequations}
for $i,j \in \{0,1\}$. $P(A_0=+1)$ is the probability of detecting a
photon in the detector $a=+1$ if the setting of $\phi_A$ was $A_0$.
They also find $25\%$ of two-photon events in which photon $1$ is
detected a time $\Delta {\cal L} /c$ before photon $2$, and $25\%$
of events in which photon $1$ is detected $\Delta {\cal L}/c$ after
photon $2$. The observers reject this $50\%$ of events and keep the
$50\%$ that are coincident. For these selected events, quantum
mechanics predicts that
\begin{equation}
P(A_i=a, B_j=b)=\frac{1}{4}\left[1+ab
\cos(\phi_{A_i}+\phi_{B_j})\right], \label{joint}
\end{equation}
where $a,b \in \{-1,+1\}$ and $\phi_{A_i}$ ($\phi_{B_j}$) is the
phase setting corresponding to $A_i$ ($B_j$).

The Bell-CHSH inequality is
\begin{equation}
-2 \le \beta_{\rm CHSH} \le 2, \label{CHSH}
\end{equation}
where
\begin{equation}
\beta_{\rm CHSH} = \langle A_0 B_0 \rangle + \langle A_0 B_1 \rangle
+ \langle A_1 B_0 \rangle - \langle A_1 B_1 \rangle.
\end{equation}
According to quantum mechanics, the maximal violation of the
Bell-CHSH inequality is $\beta_{\rm CHSH} = 2 \sqrt{2}$
\cite{Tsirelson80}, and is obtained, e.g., with $\phi_{A_0}=0$,
$\phi_{A_1}=\frac{\pi}{2}$, $\phi_{B_0}=-\frac{\pi}{4}$,
$\phi_{B_1}=\frac{\pi}{4}$.


\begin{table}[b]
\caption{\label{TableI}$32$ sets of instructions (out of $64$) of
the LHV model (the other $32$ are in Table \ref{TableII}). Each row
represents $4$ sets of local instructions (first $4$ entries) and
their corresponding contributions for the calculation of $\beta_{\rm
CHSH}$ after applying the postselection procedure of the Franson
experiment (last $4$ entries). For each row, two sets (corresponding
to $\pm$ signs) are explicitly written, while the other two can be
obtained by changing all signs.}
\begin{ruledtabular}
{\begin{tabular}{ccccccccc} $A_0$ & $A_1$ & $B_0$ & $B_1$ & $\langle
A_0 B_0 \rangle$ & $\langle A_0 B_1 \rangle$ & $\langle A_1
B_0 \rangle$ & $\langle A_1 B_1 \rangle$ \\
\hline \hline
$S+$ & $S+$ & $S+$ & $L\pm$ & $+1$ & rejected & $+1$ & rejected \\
$L+$ & $L+$ & $L+$ & $S\pm$ & $+1$ & rejected & $+1$ & rejected \\
$S+$ & $S-$ & $L\pm$ & $S+$ & rejected & $+1$ & rejected & $-1$ \\
$L+$ & $L-$ & $S\pm$ & $L+$ & rejected & $+1$ & rejected & $-1$ \\
$S+$ & $L\pm$ & $S+$ & $S+$ & $+1$ & $+1$ & rejected & rejected \\
$L+$ & $S\pm$ & $L+$ & $L+$ & $+1$ & $+1$ & rejected & rejected \\
$L\pm$ & $S+$ & $S+$ & $S-$ & rejected & rejected & $+1$ & $-1$ \\
$S\pm$ & $L+$ & $L+$ & $L-$ & rejected & rejected & $+1$ & $-1$ \\
\end{tabular}}
\end{ruledtabular}
\end{table}


\begin{table}[t]
\caption{\label{TableII}$32$ sets of instructions of the LHV model.}
\begin{ruledtabular}
{\begin{tabular}{ccccccccc} $A_0$ & $A_1$ & $B_0$ & $B_1$ & $\langle
A_0 B_0 \rangle$ & $\langle A_0 B_1 \rangle$ & $\langle A_1
B_0 \rangle$ & $\langle A_1 B_1 \rangle$ \\
\hline \hline
$S+$ & $S+$ & $S-$ & $L\pm$ & $-1$ & rejected & $-1$ & rejected \\
$L+$ & $L+$ & $L-$ & $S\pm$ & $-1$ & rejected & $-1$ & rejected \\
$S+$ & $S-$ & $L\pm$ & $S-$ & rejected & $-1$ & rejected & $+1$ \\
$L+$ & $L-$ & $S\pm$ & $L-$ & rejected & $-1$ & rejected & $+1$ \\
$S-$ & $L\pm$ & $S+$ & $S+$ & $-1$ & $-1$ & rejected & rejected \\
$L-$ & $S\pm$ & $L+$ & $L+$ & $-1$ & $-1$ & rejected & rejected \\
$L\pm$ & $S-$ & $S+$ & $S-$ & rejected & rejected & $-1$ & $+1$ \\
$S\pm$ & $L-$ & $L+$ & $L-$ & rejected & rejected & $-1$ & $+1$ \\
\end{tabular}}
\end{ruledtabular}
\end{table}


{\em LHV models for the Franson experiment.---}A LHV theory for the
Franson experiment must describe how each of the photons makes two
decisions. The $+1/-1$ decision: the decision of a detection to
occur at detector $+1$ or at detector $-1$, and the $S/L$ decision:
the decision of a detection to occur at time $t_D=t$ or a time
$t_D=t+\frac{\Delta {\cal L}}{c}$. Both decisions may be made as
late as the detection time $t_D$, and may be based on events in the
backward light cones of the detections. In a Franson-type setup both
decisions may be based on the corresponding local phase setting at
$t_D-t_I$. For a conclusive Bell test, there is no problem if
photons make the $+1/-1$ decision based on the local phase setting.
The problem is that the $50\%$ postselection procedure should be
independent on the phase settings, otherwise the Bell-CHSH
inequality (\ref{CHSH}) is not valid. In the Franson experiment the
phase setting at $t_D-t_I$ can causally affect the decision of a
detection of the corresponding photon to occur at time $t_D=t$ or a
time $t_D=t+\frac{\Delta {\cal L}}{c}$. If the $S/L$ decision can
depend on the phase settings, then, after the $50\%$ postselection
procedure, one can formally obtain not only the violations predicted
by quantum mechanics, as proven in \cite{AKLZ99}, but any value of
$\beta_{\rm CHSH}$, even those forbidden by quantum mechanics. This
is proven by constructing a family of explicit LHV models.

Consider the $64$ sets of local instructions in tables \ref{TableI}
and \ref{TableII}. For instance, if the pair of photons follows the
first set of local instructions in Table \ref{TableI}, $(A_0=)S+$,
$(A_1=)S+$, $(B_0=)S-$, $(B_1=)L+$, then, if the setting of $\phi_A$
is $A_0$ or $A_1$, photon $1$ will be detected by the detector
$a=+1$ at time $t$ (corresponding to the path $S$), and if the
setting of $\phi_B$ is $B_0$, photon $2$ will be detected by $b=-1$
at time $t$, but if the setting of $\phi_B$ is $B_1$, photon $2$
will be detected by $b=+1$ at time $t+\frac{\Delta {\cal L}}{c}$
(corresponding to the path $L$). If each of the $32$ sets of
instructions in Table \ref{TableI} occurs with probability $p/32$,
and each of the $32$ sets of instructions in Table \ref{TableII}
with probability $(1-p)/32$, then it is easy to see that, for any
value of $0 \le p \le 1$, the model gives $25\%$ of $SL$ events,
$25\%$ of $LS$ events, $50\%$ of $SS$ or $LL$ events, and satisfies
(\ref{Amarginal}) and (\ref{Bmarginal}). If $p=0$, the model gives
$\beta_{\rm CHSH}=-4$. If $p=1$, the model gives $\beta_{\rm
CHSH}=4$. If $0 < p < 1$, the model gives any value between $-4 <
\beta_{\rm CHSH} < 4$. Specifically, a maximal quantum violation
$\beta_{\rm CHSH} = 2 \sqrt{2}$, satisfying (\ref{joint}), is
obtained when $p=(2+\sqrt{2})/4$.

The reason why this LHV model is possible is that the $50\%$
postselection procedure in Franson's experiment allows the
subensemble of selected events to depend on the phase settings. For
instance, the first $8$ sets of instructions in Table \ref{TableI}
are rejected only when $\phi_B=B_1$. The main aim of this Letter is
to introduce a similar experiment which does not have this problem.

There is a previously proposed solution consisting on replacing the
beam splitters $BS_1$ and $BS_2$ in Fig.~\ref{Fig1} by switchers
synchronized with the source \cite{BGTZ99}. However, these active
switchers are replaced in actual experiments by passive beam
splitters \cite{TBZG00, BGTZ99} that force a Franson-type
postselection with the same problem described above.

One way to avoid the problem is to make an extra assumption, namely
that the decision of being detected at time $t_D=t$ or a time
$t_D=t+\frac{\Delta {\cal L}}{c}$ is actually made at the first beam
splitter, before having information of the local phase settings
\cite{AKLZ99, Franson99}. This assumption is similar to the fair
sampling assumption, namely that the probability of rejection does
not depend on the measurement settings. As we have seen, there are
local models that do not satisfy this assumption. The experiment we
propose does not require this extra assumption.


\begin{figure}[htb]
\centerline{\includegraphics[width=1.04 \columnwidth]{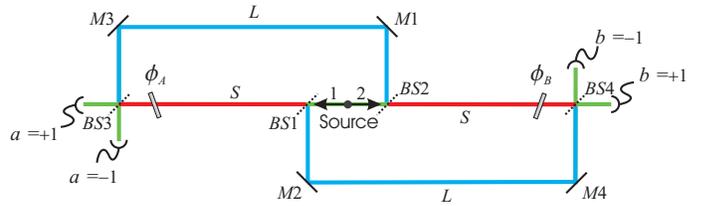}}
\caption{\label{Fig2} The generic setup of the proposed energy-time
(and time-bin) Bell experiment.}
\end{figure}


{\em Proposed energy-time entanglement Bell experiment.---}The setup
of the new Bell experiment is illustrated in Fig.~\ref{Fig2}. The
source emits two photons, photon $1$ to the left and photon $2$ to
the right. The $S$ path of photon $1$ (photon $2$) ends on the
detectors $a$ on the left ($b$ on the right). The difference with
Fig.~\ref{Fig1} is that now the $L$ path of photon $1$ (photon $2$)
ends on the detectors $b$ ($a$). In this setup, the two photons end
in different sides only when both are detected in coincidence. If
one photon takes $S$ and the other photon takes $L$, both will end
on detectors of the same side. An interferometer with this last
property is described in \cite{RVDM08}.

The data that the observers must record is the same as in Franson's
experiment. The setup must satisfy the following requirements: (I')
To have two-photon interference, the emission of the two photons
must be simultaneous, the moment of emission unpredictable, and both
arms of the setup identical. The phase stabilization of the entire
setup of Fig.~\ref{Fig2} is more difficult than in Franson's
experiment. (II') Single-photon interference is not possible in the
setup of Fig.~\ref{Fig2}. (III') To temporally distinguish two
photons arriving at the same detector at times $t$ and
$t+\frac{\Delta {\cal L}'}{c}$, where $\Delta {\cal L}'=2 [d({\rm
Source},BS2)+d(BS2,M1)]$ (see Fig.~\ref{Fig2}), the dead time of the
detectors must be smaller than $\frac{\Delta {\cal L}'}{c}$. For
detectors with a dead time of $1$ ns, ${\Delta {\cal L}'} > 30$ cm.
(IV') The probability of two two-photons events in $\frac{\Delta
{\cal L}'}{c}$ must be negligible. This naturally occurs when using
standard non-linear crystals pumped continuously. (V') To prevent
that the local phase setting at one side can affect the outcome at
the other side, the local phase settings must randomly switch
($\phi_A$ between $A_0$ and $A_1$, and $\phi_B$ between $B_0$ and
$B_1$) with a frequency of the order $c/D'$, where $D'=d({\rm
Source},\phi_A)\gg \Delta {\cal L}'$.

There is a trade-off between the phase stabilization of the
apparatus (which requires a short interferometer) and the prevention
of reciprocal influences between the two local phase settings (which
requires a long interferometer). By considering a random phase
modulation frequency of 300 kHz, an interferometer about 1 km long
would be needed. Current technology allows us to stabilize
interferometers of up 4 km long (for instance, one of the
interferometers of the LIGO experiment is 4 km long). With these
stable interferometers, the experiment would be feasible.

The predictions of quantum mechanics for the setup of
Fig.~\ref{Fig2} are similar to those in Franson's proposal: Eqs.
(\ref{Amarginal}) and (\ref{Bmarginal}) hold, there is $25\%$ of
events in which both photons are detected on the left at times $t$
and $t+\frac{\Delta {\cal L}'}{c}$, $25\%$ of events in which both
photons are detected on the right, and $50\%$ of coincident events
for which (\ref{joint}) holds. The observers must keep the
coincident events and reject those giving two detections on
detectors of the same side. The main advantages of this setup are:
(i) The rejection of events is local and does not require
communication between the observers. (ii) The selection and
rejection of events is independent of the local phase settings. This
is the crucial difference with Franson's experiment and deserves a
detailed examination. First consider a selected event: both photons
have been detected at time $t_D$, one in a detector $a$ on the left,
and the other in a detector $b$ on the right. $t_I$ is the time a
photon takes from $\phi_A$ ($\phi_B$) to a detector $a$ ($b$). The
phase setting of $\phi_A$ ($\phi_B$) at $t_D-t_I$ is in the backward
light cone of the photon detected in $a$ ($b$), but the point is,
could a different value of one or both of the phase settings have
caused that this selected event would become a rejected event in
which both photons are detected on the same side? The answer is no.
This would require a mechanism to make one detection to ``wait''
until the information about the setting in other side comes.
However, when this information has finally arrived, the phase
settings (both of them) have changed, so this information is useless
to base a decision on it.

Now consider a rejected event. For instance, one in which both
photons are detected in the detectors $a$ on the left, one at time
$t_D=t$, and the other at $t_D=t+\frac{\Delta {\cal L}'}{c}$. Then,
the phase settings of $\phi_B$ at times $t_D-t_I$ are out of the
backward light cones of the detected photons. The photons cannot
have based their decisions on the phase settings of $\phi_B$. A
different value of $\phi_A$ cannot have caused that this rejected
event would become a selected event. This would require a mechanism
to make one detection to wait until the information about the
setting arrives to the other side, and when this information has
arrived, the phase setting of $\phi_A$ has changed so this
information is useless.

For the proposed setup, there is no physical mechanism preserving
locality which can turn a selected (rejected) event into a rejected
(selected) event. The selected events are independent of the local
phase settings. For the selected events, only the $+1/-1$ decision
can depend on the phase settings. This is exactly the assumption
under which the Bell-CHSH inequality (\ref{CHSH}) is valid.
Therefore, an experimental violation of (\ref{CHSH}) using the setup
of Fig.~\ref{Fig2} and the postselection procedure described before
provides a conclusive (assuming perfect detectors) test of local
realism using energy-time (or time-bin) entanglement. Indeed, the
proposed setup opens up the possibility of using genuine energy-time
or time-bin entanglement for many other quantum information
experiments.


The authors thank J.D. Franson, J.-\AA. Larsson, T. Rudolph, and M.
\.{Z}ukowski for their comments. This work was supported by Junta de
Andaluc\'{\i}a Excellence Project No. P06-FQM-02243 and by
Finanziamento Ateneo 07 Sapienza Universit\'{a} di Roma.


\end{document}